\documentclass[pra,aps,twocolumn,showpacs,superscriptaddress]{revtex4}
\usepackage{epsfig}

\begin{document}

\title{Entanglement conditions for two-mode states}
\author{Mark Hillery}
\affiliation{Department of Physics, Hunter College of CUNY, 695 Park Avenue,
New York, NY 10021}
\author{M. Suhail Zubairy}
\affiliation{Institute for Quantum Studies and Department of Physics,
Texas A\&M University, College Station, TX 77843}
\date{\today}

\begin{abstract}
We provide a class of inequalities whose violation shows the presence of
entanglement in two-mode systems.  We
initially consider observables that are quadratic in the mode creation and 
annihilation operators and find conditions under which a two-mode state is 
entangled. Further examination allows us to formulate additional conditions 
for detecting entanglement. We conclude by showing how the methods used here 
can be extended to find entanglement in systems of more than two modes.
\end{abstract}

\maketitle

Entanglement has proven to be a valuable resource in quantum information 
processing. However, determining whether or not a state is entangled is often 
far from simple.  Methods such as the Peres-Horodecki positive partial transpose
condition \cite{horodecki}, entanglement witnesses \cite{kraus}, and 
hierarchies of entanglement conditions \cite{doherty} exist, but are not always 
straightforward to apply.
In particular, for systems with continuous degrees of 
freedom, such as particle position or momentum or the quadrature components of
field modes, the number of available criteria for detecting entanglement 
is very limited. Each of the known criteria detects only a subset of the
set of entangled states.  In many cases, these criteria are in the form of 
inequalities \cite{simon1}-\cite{toth}. In general they provide only 
sufficient conditions for detecting entanglement \cite{zubairy}. 
The utility of most of these inequalities is however limited for non-Gaussian 
bipartite states. For example, none of these conditions can detect the fact 
that the state $(|0\rangle_{a}|1\rangle_{b} +|1\rangle_{a}|0\rangle_{b})/
\sqrt{2}$ is an entangled state, though it should be pointed out that it can be shown
to be entangled by the application of other entanglement tests \cite{horodecki,kim}.  
This indicates that there is a need to find additional simple, and ideally, experimentally
accessible conditions that can establish  whether a state is entangled.

In this paper we provide a class of inequalities for detecting entanglement. These inequalities
arise from examining uncertainty relations.  The use of uncertainty relations to establish 
conditions for detecting entanglement has been pursued by Hofmann and Takeuchi 
\cite{hofmann} and by Guehne \cite{guehne}.  We begin by examining observables that are quadratic in the mode creation and annihilation operators.
These observables were used previously to define sum and difference squeezing,
forms of higher-order squeezing \cite{hillery}.  
These quantities and their uncertainties are, in principle,
measurable, so that the conditions we derive could be used 
in a laboratory to detect entanglement.  
We find that the conditions formulated in terms of these variables lead to a
host of other conditions for detecting entanglement. Finally, we shall
briefly discuss how some
of these conditions can be extended to detect entanglement in
systems consisting of more than two modes.

Consider two modes of the electromagnetic field, where 
$a$ and $a^{\dagger}$ are the 
annihilation and creation operators of the first mode and $b$ and
$b^{\dagger}$ are the 
annihilation and creation operators of the second.  We define the operators
$L_{1}  =  ab^{\dagger}+a^{\dagger}b$ and   
$L_{2}  =  i(ab^{\dagger} - a^{\dagger}b)$ .
Operators, proportional to these, along with one proportional to
the operator $L_{3}=a^{\dagger}a+b^{\dagger}b$, 
form a representation of
the su(2) Lie algebra, i.e., $J_{i}=L_{i}/2$ ($i=1-3$) satisfy the commutation 
relations $[J_{k},J_{m}]=i\epsilon_{kmn}J_{n}$.  Entanglement conditions 
expressed in terms of angular momentum
operators have been derived by a number of authors
\cite{sorensen1}-\cite{simon2}.
It follows, on calculating the uncertainties of these variables and adding 
them, that 
\begin{eqnarray}
(\Delta L_{1})^{2}+(\Delta L_{2})^{2} &=& 2(\langle (N_{a}+1)N_{b}\rangle 
+ \langle N_{a}(N_{b}+1)\rangle \nonumber \\
&&-2|\langle ab^{\dagger}\rangle |^{2}) ,
\end{eqnarray}
where $N_{a}=a^{\dagger}a$ and $N_{b}=b^{\dagger}b$.  Now suppose that the 
state we are
considering is a product of a state in the $a$ mode and another state 
in the $b$ mode.  Then
the expectation values in the above expression factorize into 
products of $a$-mode and
$b$-mode expectation values.  We then have that
\begin{eqnarray}
(\Delta L_{1})^{2}+(\Delta L_{2})^{2} &=& 2(\langle (N_{a}+1)\rangle 
\langle N_{b}\rangle + 
\langle N_{a}\rangle \langle (N_{b}+1)\rangle \nonumber \\
&&-2|\langle a\rangle \langle b^{\dagger}\rangle |^{2}) ,
\end{eqnarray}
Noting that the Schwarz inequality implies that $|\langle a \rangle |^{2} 
\leq \langle N_{a}\rangle$ and $ |\langle b \rangle |^{2} \leq 
\langle N_{b}\rangle$,
we find that for a product state
\begin{equation}
\label{prod1}
(\Delta L_{1})^{2}+(\Delta L_{2})^{2} \geq 2(\langle N_{a}\rangle + 
\langle N_{b}\rangle ) .
\end{equation}

This inequality can be extended to any separable state by using a result 
of Hofmann and Takeuchi
\cite{hofmann}.  For a density matrix $\rho = \sum_{m}p_{m}\rho_{m}$ and
a variable $S$, 
we have that
\begin{equation}
\label{decomp}
(\Delta S)^{2} \geq \sum_{m} p_{m} (\Delta S_{m})^{2} ,
\end{equation}
where $(\Delta S_{m})^{2} $ is the uncertainty of $S$ calculated in 
the state $\rho_{m}$.  If the
original state $\rho$ is separable, then all of the states $\rho_{m}$ 
can be taken to be product
states for which the inequality in Eq.\ (\ref{prod1}) holds.  
Then, Eq.\ (\ref{decomp}) implies that
Eq.\ (\ref{prod1}) holds for the state $\rho$ as well.  Hence, 
Eq.\ (\ref{prod1}) is valid for any
separable state. It can be easily shown that Eq.\ (\ref{prod1}) is 
violated for the Bell state $|\psi_{01}\rangle = (|0\rangle_{a} |1\rangle_{b} 
+ |1\rangle_{a} |0\rangle_{b})/\sqrt{2}$ .

We can gain more insight if we consider the uncertainty relation obeyed 
by $L_{1}$ and $L_{2}$,
\begin{equation}
(\Delta L_{1})(\Delta L_{2})\geq  |\langle N_{a}-N_{b}\rangle | .
\end{equation}
This implies that
\begin{equation}
\label{unc1}
(\Delta L_{1})^{2}+(\Delta L_{2})^{2} \geq 2 | \langle N_{a}-N_{b}\rangle | .
\end{equation}
Comparing this result, which holds for any state, to Eq.\ (\ref{prod1}), 
which holds for separable
states, we see that the right-hand side of Eq.\ (\ref{unc1}) 
is always less than or equal to that
of Eq.\ (\ref{prod1}).  Consequently, there are states that violate
Eq.\ (\ref{prod1}) while satisfying
Eq.\ (\ref{unc1}), and the state in the previous paragraph is an example 
of such a state.

It is also worthwhile to see how the condition in Eq.\ (\ref{prod1})performs 
for a mixed state.  Consider the state
\begin{equation}
\rho = s|\psi_{01}\rangle\langle \psi_{01}| +\frac{1-s}{4}P_{01}  ,
\end{equation}
where $0\leq s \leq 1$ and $P_{01}$ is the projection operator onto thespace
spanned by the vectors 
$\{ |0\rangle_{a} |0\rangle_{b} , |0\rangle_{a} |1\rangle_{b} , |1\rangle_{a}
|0\rangle_{b} ,|1\rangle_{a} |1\rangle_{b}  \}$.  
We find that $(\Delta L_{1})^{2}+(\Delta L_{2})^{2}  =  3-s-s^{2}$ and
$2(\langle N_{a}\rangle + \langle N_{b}\rangle )  =  2$  ,
so that violation of the inequality in Eq.\ (\ref{prod1}) shows that 
the state is entangled if $s^{2}+s-1>0$, or $1\geq s > (\sqrt{5}-1)/2$.

An examination of the condition in Eq.\ (\ref{prod1}) shows us that 
the state is entangled if
\begin{equation}
\label{prod2}
\langle N_{a}N_{b}\rangle < |\langle ab^{\dagger}\rangle |^{2}  .
\end{equation}
Note that the Schwarz inequality implies that
\begin{equation}
 |\langle ab^{\dagger}\rangle |^{2} \leq \langle N_{a}(N_{b}+1) \rangle  ,
\end{equation}
so there are states that can satisfy the inequality in Eq.\ (\ref{prod2}).  
This condition suggests
that there is a family of similar conditions for detecting entanglement, 
where instead of considering
the operator $ab^{\dagger}$ we consider instead $a^{m}(b^{\dagger})^{n}$. 
 For a pure product state we have that
\begin{equation}
\label{prod3}
|\langle a^{m}(b^{\dagger})^{n}\rangle |^{2} =|\langle a^{m}\rangle |^{2} 
|\langle b^{n}\rangle |^{2}
\leq \langle (a^{\dagger})^{m}a^{m}\rangle \langle (b^{\dagger})^{n}
b^{n}\rangle  ,
\end{equation}
or, because for a product state $ \langle (a^{\dagger})^{m}a^{m}\rangle 
\langle (b^{\dagger})^{n}b^{n}\rangle =  \langle (a^{\dagger})^{m}a^{m} 
(b^{\dagger})^{n}b^{n}\rangle$,
it is also true that
\begin{equation}
\label{prod4}
|\langle a^{m}(b^{\dagger})^{n}\rangle |^{2} \leq  \langle(a^{\dagger})^{m}
a^{m}(b^{\dagger})^{n}b^{n}\rangle  .
\end{equation}
It is this relation that will lead to a generalization of the entanglement 
condition in Eq.\ (\ref{prod2}), but before it does, we need to show that 
it holds for any separable state and not just for product states. 
Consider the density matrix for a general separable state given by
$\rho =\sum_{k}p_{k}\rho_{k}$, where $\rho_{k}$ is a density matrix
corresponding to a pure product state, and $p_{k}$ is the probability of 
$\rho_{k}$.  The probabilities satisfy the condition $\sum_{k}p_{k}=1$.  
Defining $A=a^{m}$ and $B=b^{n}$, we have that
\begin{eqnarray}
|\langle AB^{\dagger}\rangle | & \leq & \sum_{k} p_{k} |{\rm Tr}
(\rho_{k}AB^{\dagger})| 
\nonumber  \\
 & \leq & \sum_{k}p_{k}( \langle A^{\dagger}AB^{\dagger}B\rangle_{k})^{1/2} ,
\end{eqnarray}
where  $\langle A^{\dagger}AB^{\dagger}B\rangle_{k}={\rm Tr}(\rho_{k}  
A^{\dagger}AB^{\dagger}B)$.  We can now apply the Schwarz inequality 
to obtain
\begin{eqnarray}
|\langle AB^{\dagger}\rangle |&  \leq & \left( \sum_{k}p_{k} \right)^{1/2} 
\left( \sum_{k}p_{k}\langle A^{\dagger}AB^{\dagger}B\rangle_{k} \right)^{1/2} 
\nonumber \\
 & \leq & (\langle A^{\dagger}AB^{\dagger}B\rangle )^{1/2} ,
\end{eqnarray}
which shows that the inequality in Eq.\ (\ref{prod4}) does indeed hold for 
all separable states. Therefore, we can conclude that a state is entangled if
\begin{equation}
\label{ent1}
|\langle a^{m}(b^{\dagger})^{n}\rangle |^{2}  > 
\langle(a^{\dagger})^{m}a^{m}(b^{\dagger})^{n}b^{n}\rangle  .
\end{equation}

Let us now turn our attention to the variables $K_{1}=ab 
+ a^{\dagger}b^{\dagger}$ and $K_{2}=i( a^{\dagger}b^{\dagger}-ab )$.
One half times these operators along with one half times the operator
the operator $K_{3}=a^{\dagger}a -b^{\dagger}b$ form a representation of the 
su(1,1) Lie algebra.  As before, we would like to find inequalities 
involving these variables that tell us whether a two-mode state is 
entangled or not.  The strategy that we employed before, adding
the uncertainties and assuming the expectation values can be factorized,
leads to the inequality for product states
$(\Delta K_{1})^{2}+(\Delta K_{2})^{2} \geq 2(\langle N_{a}\rangle 
+ \langle N_{b}\rangle +1)$ .
However, if we employ the uncertainty relation, $\Delta K_{1} \Delta K_{2} 
\geq \langle N_{a}+N_{b}+1\rangle$, we find that the above inequality holds 
for all states, and is therefore not useful
for determining whether a state is entangled or not.

We can obtain something useful if we pursue a different path.  The guiding 
idea is that the 
``eigenstates'' (the reason for the quotation marks is that these states 
are, in general, not 
normalizable, and hence do not lie in the Hilbert space of 
two-mode states) of operators such 
as $K_{1}$ and $K_{2}$ are highly entangled.  States whose 
uncertainty in one of these variables is small
will be close to one of these eigenstates, and will also be entangled.  
Therefore, for a state to be separable, its uncertainty in one of these 
variables must be greater than some lower bound.  What we shall show is 
that in the case of  $K_{1}$ and $K_{2}$, that lower bound is $1$.

In order to make the discussion more general, define the variable
\begin{equation}
K(\phi )=  e^{i\phi} a^{\dagger}b^{\dagger} + e^{-i\phi}ab .
\end{equation}
Note that $K(0)=K_{1}$ and $K(\pi /2)=K_{2}$.  We then have that
\begin{eqnarray}
(\Delta K(\phi ))^{2} & = & \langle (a^{\dagger}b^{\dagger}
-\langle a^{\dagger}b^{\dagger}\rangle )
(ab-\langle ab\rangle )\rangle \nonumber \\
&&+\langle (ab-\langle ab\rangle )(a^{\dagger}b^{\dagger}
-\langle a^{\dagger}b^{\dagger} \rangle )\rangle \nonumber \\
&&+ e^{2i\phi } \langle (a^{\dagger}b^{\dagger}
-\langle a^{\dagger}b^{\dagger} \rangle )^{2}\rangle \nonumber \\
&&+ e^{-2i\phi } \langle (ab-\langle ab\rangle )^{2} \rangle .
\end{eqnarray}
We again employ the Schwarz inequality to give us
\begin{eqnarray}
|\langle (ab - \langle ab\rangle )^{2} \rangle | & \leq & 
[ \langle (ab-\langle ab\rangle ) 
(a^{\dagger}b^{\dagger} -\langle a^{\dagger}b^{\dagger} \rangle )\rangle 
\nonumber  \\
 & & \langle (a^{\dagger}b^{\dagger}-\langle a^{\dagger }b^{\dagger}\rangle )
(ab-\langle ab\rangle )\rangle ]^{1/2} .
\end{eqnarray}
This gives us that 
\begin{eqnarray}
(\Delta K(\phi ))^{2} &\geq & [ \langle (ab-\langle ab\rangle ) 
(a^{\dagger}b^{\dagger} -\langle a^{\dagger}b^{\dagger} \rangle )\rangle^{1/2}
\nonumber  \\
 & & -\langle (a^{\dagger}b^{\dagger}-\langle a^{\dagger }b^{\dagger}\rangle ) 
(ab-\langle ab\rangle )\rangle^{1/2} ]^{2}  \nonumber \\
 & \geq & [( \langle (N_{a}+1)(N_{b}+1)\rangle
-|\langle ab\rangle |^{2})^{1/2} \nonumber \\
 & & -( \langle N_{a}N_{b}\rangle -|\langle ab\rangle |^{2})^{1/2}]^{2}  .
\end{eqnarray}

This inequality is valid for all states, but if the state is a product state 
this becomes
\begin{eqnarray}
(\Delta K(\phi ))^{2} & \geq & [( \langle (N_{a}+1)\rangle 
\langle (N_{b}+1)\rangle -|\langle a\rangle
\langle b\rangle |^{2})^{1/2}  \nonumber  \\   
 & & -( \langle N_{a}\rangle \langle N_{b}\rangle 
-|\langle a\rangle \langle b\rangle |^{2})^{1/2}]^{2}  .
\end{eqnarray}

Now let us examine the quantity on the right-hand side of the above 
inequality.  Setting $x=\langle N_{a}\rangle $, $y= \langle N_{b}\rangle $, 
and $z=|\langle ab \rangle |^{2}$, we want to find the minimum of the function
\begin{equation}
F(x,y)=\sqrt{(x+1)(y+1)-z} - \sqrt{xy-z} ,
\end{equation}
in the region $xy\geq z \geq 0$.  By setting $\partial F/\partial x$ and 
$\partial F/\partial y$ equal to zero, we find that $F(x,y)$ has no local 
maxima or minima in the region of interest, so that the
minimum of the function must lie on the boundary.  This means we have to 
look at how $F$ behaves on the curve $xy=z$ and as $x$ and $y$ go to infinity.
On the curve $xy=z$ we find that
\begin{equation}
F(x,z/x)=\left( x + \frac{z}{x}+ 1 \right)^{1/2} \geq 1   .
\end{equation}
Now let us consider what happens as $x,y \rightarrow \infty$.  We first 
note that
\begin{equation}
F(x,y)=\int_{xy-z}^{(x+1)(y+1)-z} du \frac{1}{2\sqrt{u}} \geq 
\frac{x+y+1}{2\sqrt{(x+1)(y+1)-z}}  .
\end{equation}
Continuing, we find
\begin{eqnarray}
F(x,y) & \geq &  \frac{x+y+1}{2\sqrt{(x+1)(y+1)}}= \frac{(x+1)+(y+1)-1}
{2\sqrt{(x+1)(y+1)}} \nonumber  \\
 & \geq & \frac{1}{2}\left[ \sqrt{\frac{x+1}{y+1}} + \sqrt{\frac{y+1}{x+1}} 
-\frac{1}{\sqrt{(x+1)(y+1)}} \right]  \nonumber \\
 & \geq &  1- \frac{1}{2\sqrt{(x+1)(y+1)}} .
\end{eqnarray}
Therefore, we can conclude that as $x,y \rightarrow \infty$ we have that 
$F(x,y)\geq 1$.  Finally,
this gives us $(\Delta K(\phi )) \geq 1$ for a product state, and the 
argument in \cite{hofmann} (see Eq.\ (\ref{decomp}) ) then implies that it 
is true for any separable state.  Consequently, if for some state 
$(\Delta K(\phi )) < 1$, we can conclude it is entangled.

Both $K_{1}$ and $K_{2}$ are measurable.  If the two modes are sent into a 
nonlinear crystal,
to lowest order in the nonlinearity, the quadrature components of the 
mode corresponding to
their sum frequency are proportional to $K_{1}$ and $K_{2}$ \cite{hillery}.  
These quadratures can then be determined by means of homodyne measurements.

Let us exhibit a state for which $\Delta K_{1}<1$.  Consider the state
\begin{equation}
|\psi\rangle = \frac{1}{\sqrt{\eta}}\sum_{n=0}^{\infty}(-1)^{n}
\frac{x^{n}}{\sqrt{2n+1}} |2n\rangle_{a} |2n\rangle_{b} ,
\end{equation}
where $0 \leq x <1$ and 
\begin{equation}
\eta = \frac{1}{2x}\ln \left( \frac{1+x}{1-x} \right) .
\end{equation}
For this state we find that $\langle ab \rangle =0$, and 
\begin{eqnarray}
\langle \psi |a^{2}b^{2}|\psi \rangle & = & -\frac{1}{\eta}\sum_{n=0}^{\infty}
\frac{(2n+2)(2n+1)}
{[(2n+3)(2n+1)]^{1/2}} x^{2n+1}  \nonumber \\
\langle\psi | (N_{a}+1)(N_{b}+1)|\psi\rangle & = & \frac{1+x^{2}}
{\eta(1-x^{2})^{2}} \nonumber \\
\langle \psi |N_{a}N_{b}|\psi \rangle & = & \frac{3x^{2}-1}
{\eta (1-x^{2})^{2}} + 1 ,
\end{eqnarray}
which implies
\begin{equation}
(\Delta K_{1})^{2} =1 + \frac{4x^{2}}{\eta (1-x^{2})^{2}} 
-  \frac{2}{\eta}\sum_{n=0}^{\infty} 
\frac{(2n+2)(2n+1)} {[(2n+3)(2n+1)]^{1/2}} x^{2n+1} .
\end{equation}
We can find a lower bound for the sum in the above equation, which gives us an
upper bound for $(\Delta K_{1})^{2}$.   We obtain
\begin{equation}
(\Delta K_{1})^{2} \leq 1 + \frac{4}{\eta}\left[ -\frac{1}{\sqrt{3}} x 
+\frac{x^{2}[1-x(2-x^{2})]}{(1-x^{2})^{2}} \right]  .
\end{equation}
Noting that near $x=0$ we have that $\eta$ is approximately equal 
to $1+(x^{2}/3)$ we find that
near $x=0$ the right-hand side of the above equation behaves 
like $1-(4x/\sqrt{3})$, so that $\Delta K_{1}$ for this state can indeed 
be less than $1$.

In analogy to what was done for the su(2) variables it is possible 
to find other relations that must be obeyed by separable states.  For example,
in the case of product states we have that
\begin{equation}
\label{ent2}
|\langle ab\rangle |=|\langle a\rangle \langle b\rangle | \leq 
[\langle N_{a}\rangle \langle N_{b}\rangle ]^{1/2} ,
\end{equation}
and what we shall now do is show that this inequality is obeyed by all 
separable states. Therefore,
a violation of this inequality implies that the state is entangled.  
In fact, we shall show that for
any positive integers $m$ and $n$, a separable state must satisfy the 
condition
\begin{equation}
\label{ent3}
|\langle a^{m}b^{n}\rangle | \leq [\langle (a^{\dagger})^{m}a^{m}\rangle 
\langle (b^{\dagger})^{n}b^{n}\rangle ]^{1/2}  .
\end{equation}
Clearly Eq.\ (\ref{ent2}) is a special case of Eq.\ (\ref{ent3}).

As before, consider the density matrix of a general separable state 
$\rho =\sum_{k}p_{k}\rho_{k}$,
where $\rho_{k}$ is a density matrix corresponding to a pure product state, 
and $p_{k}$ is the probability of $\rho_{k}$. 
Again, setting $A=a^{m}$ and $B=b^{n}$, we have that
\begin{eqnarray}
| \langle AB \rangle |^{2} & \leq & \sum_{k,l}p_{k}p_{l}|{\rm Tr}(\rho_{k}AB)|
|{\rm Tr}(\rho_{l}B^{\dagger}A^{\dagger})|  \nonumber \\
 &\leq & \sum_{k,l}p_{k}p_{l}(\langle A^{\dagger}A\rangle_{k} \langle 
B^{\dagger}B\rangle_{k}
\langle A^{\dagger}A\rangle_{l} \langle B^{\dagger}B\rangle_{l})^{1/2}  .
\end{eqnarray}
In terms of the quantities $\langle A^{\dagger}A\rangle_{k} = {\rm Tr}(A^{\dagger}A\rho_{k}) = x_{k}$ and  
$\langle B^{\dagger}B\rangle_{k} = {\rm Tr}(B^{\dagger}B\rho_{k}) = y_{k}$,
this inequality can be rewritten as
\begin{equation}
| \langle AB^{\dagger} \rangle |^{2}  \leq  \sum_{k} p_{k}^{2}x_{k}y_{k} 
+2\sum_{k>l}p_{k}p_{l} (x_{k}y_{k}x_{l}y_{l})^{1/2} .
\end{equation}
Next we consider $\langle A^{\dagger}A\rangle \langle B^{\dagger}B\rangle =  
\sum_{k} p_{k}^{2}x_{k}y_{k} 
+ \sum_{k>l}p_{k}p_{l}(x_{k}y_{l}+x_{l}y_{k})$.
As $x_{k}y_{l}+x_{l}y_{k} \geq 2(x_{k}y_{k}x_{l}y_{l})^{1/2}$,
we see that the inequality in Eq.\ (\ref{ent3}) holds for all separable 
states, i.e., if a state
violates this inequality, it must be entangled.

Returning to the case $m=n=1$, we have that for a general state $|\langle ab\rangle | \leq  [\langle N_{a}+ 1\rangle 
\langle N_{b} \rangle ]^{1/2}$, which suggests that there are states that do violate the inequality in 
Eq.\ (\ref{ent2}).  An example of one that does is the two-mode squeezed 
vacuum state
\begin{equation}
|\psi\rangle = (1-x^{2})^{1/2}\sum_{n=0}^{\infty} x^{n}|n\rangle_{a} 
|n\rangle_{b} ,
\end{equation}
where $0\leq x \leq 1$.  For this state we find that $ [\langle N_{a}\rangle 
\langle N_{b} \rangle ]^{1/2} =x^{2}/(1-x^{2})<x/(1-x^{2})
=|\langle ab \rangle |$,
so that we conclude from Eq.\ (\ref{ent2}) that this state is entangled.

We have derived a family of entanglement conditions for two-mode states.  
They enlarge the set of states that can be shown to be entangled by means of relatively simple
conditions.  Some of these conditions provide, in principle, measurable tests of entanglement, 
that is, all of the quantities appearing in the inequalities can be measured in the laboratory.

In closing, we point out that the methods employed here are not confined to 
demonstrating entanglement in two-mode states.  To show this we briefly 
consider a three-mode example.  A more thorough analysis will be left to 
future work.  Consider three modes whose annihilation operators are $a$, $b$, and $c$.  
For a state that is
a tensor product of individual states for each of the modes, we have that 
$|\langle ab^{\dagger}c^{\dagger}\rangle |  =  |\langle a \rangle 
\langle b\rangle \langle c \rangle | \leq  (\langle N_{a} \rangle 
\langle N_{b}\rangle \langle N_{c} \rangle )^{1/2} =
\langle N_{a}N_{b}N_{c}\rangle^{1/2}$.
For a state that is completely separable in the three modes, that is one 
that can be expressed as
$\rho = \sum_{k}p_{k}\rho_{ak}\otimes\rho_{bk}\otimes\rho_{ck}$,
we find that
$|\langle ab^{\dagger}c^{\dagger}\rangle |  = \sum_{k}p_{k} |\langle a \rangle_{k} 
\langle b\rangle_{k} \langle c \rangle_{k} |$, which implies that
\begin{eqnarray}
|\langle ab^{\dagger}c^{\dagger}\rangle | & \leq &  \sum_{k}p_{k}  
(\langle N_{a}N_{b}N_{c} \rangle_{k})^{1/2}  \nonumber \\
& \leq & (\sum_{k}p_{k})^{1/2}(\sum_{k}p_{k}\langle N_{a}N_{b}N_{c} \rangle_{k})^{1/2}
\nonumber \\
 & \leq &  \langle N_{a}N_{b}N_{c}\rangle^{1/2},
\end{eqnarray}
where the next to last step follows from the Schwarz inequality.  If a 
state is completely separable, it must obey this inequality, and, therefore, 
if the inequality is violated, the state
will be entangled.  An example of a state that does violate this inequality 
is given by $|\psi\rangle =( |1\rangle_{a}|0\rangle_{b}|0\rangle_{c} 
+ |0\rangle_{a}|1\rangle_{b}|1\rangle_{c} )/\sqrt{2}$,
which is a kind of GHZ state.
In particular, for this state $\langle N_{a}N_{b}N_{c}\rangle =0$, and 
$|\langle ab^{\dagger}c^{\dagger}\rangle |=1/2$, which clearly violates 
the inequality $|\langle ab^{\dagger}c^{\dagger}\rangle |  
\leq   \langle N_{a}N_{b}N_{c}\rangle^{1/2}$.
Therefore, we see that the types of inequalities developed here can be 
extended to study the multipartite entanglement of continuous-variable systems.

{\bf Note added}:  After submission of this paper, publications on very similar topics by Agarwal
and Biswas \cite{agarwal} and Shchukin and Vogel \cite{vogel} have appeared.

We would like to thank Vladimir Bu\v{z}ek for useful comments.
This research is supported by the Air Force Office of Scientific Research, 
DARPA-QuIST, and the TAMU Telecommunication and Informatics Task 
Force (TITF) initiative.

\end{document}